| | |
|---|---|
| Article title: | Fine-tuning the H-scan for visualizing types of tissue scatterers |
| Authors: | Kevin J. Parker[1,a)] Jihye Baek[1] |
| | Department of Electrical & Computer Engineering, University of Rochester, Rochester, New York 14627, USA |
| Date: | 12/10/2019 |
| Running title: | Fine-tuning the H-scan |

[a)] Author to whom correspondence should be addressed. Electronic mail: kevin.parker@rochester.edu

**ABSTRACT**

The H-scan approach is a matched filter methodology that aims to add information to the traditional ultrasound B-scan. The theory is based on the differences in the echoes produced by different classes of reflectors or scatterers. Matched filters can be created for different types of scatterers, whereby the maximum output indicates a match and color schemes can be used to indicate the class of scatterer responsible for echoes, providing a visual interpretation of the results. However, within the theory of weak scattering from a variety of shapes, small changes in the size of the inhomogeneous objects will create shifts in the scattering transfer function. In this paper, we determine the shift in center frequency, along with time and frequency domain characteristics of echoes and argue for a general power law transfer function to capture the major characteristics of scattering types. With this general approach, the H-scan matched filters can be set to elicit more fine grain shifts in scattering types, commensurate with more subtle changes in tissue morphology. Compensation for frequency dependent attenuation is helpful for avoiding beam softening effects with increasing depths. Examples from phantoms and normal and pathological tissues are provided to demonstrate that the H-scan analysis and displays are sensitive to scatterer size and morphology, and can be adapted to conventional imaging systems.

## I. INTRODUCTION

The mathematical treatment of scattering of light and sound has a rich history spanning over 100 years. The scattering of ultrasound from tissues forms the basis for the worldwide use of ultrasound imaging for diagnostic purposes, and an uncountable number of these images are obtained every day. It is generally understood that inhomogeneities within tissues, specifically localized changes in acoustic impedance or density and compressibility, are responsible for the echoes that are captured by imaging systems. A longstanding goal within the research community has been to supplement the traditional B-scan image of tissue with additional quantitative information about the scatterers, linked to the structure or size or statistical properties of the underlying tissue and cellular structures (Lizzi *et al.*, 1983; Mamou and Oelze, 2013). However, recently, a new hypothesis has been formulated that the cylindrical-shaped fluid vessels within tissue parenchyma are predominantly responsible for the echoes commonly captured by imaging systems (Parker *et al.*, 2019; Parker, 2019b, a). Within this framework and within a more general framework encompassing different classical scattering models, the H-scan analysis can be helpful in discriminating between tissue types (Parker, 2016c, a; Khairalseed *et al.*, 2017; Ge *et al.*, 2018; Khairalseed *et al.*, 2019a; 2019b). In essence, the H-scan seeks to make a matched filter for the case of a bandpass ultrasound pulse incident on a tissue scatterer or reflector, and producing an echo. The output of a set of matched filters covering the different types of expected scatterers, shown in Figure 1, can be examined to determine a best fit to any particular echo or segment, and this additional information can be encoded as a color overlay or a quantitative map. In this way the H-scan analysis provides relatively high resolution and localized information about the nature of the underlying tissue.

In this article we make a series of arguments leading to the fine-tuning of the H-scan analysis for more sensitive discrimination of small changes in scatterers within different regions or over time in response to the progression of disease or the effects of therapy. First, under the hypothesis that the fluid-filled cylindrical vessels and channels form the dominant scattering sites, an examination of theory concludes that small shifts in the center frequency of a bandpass interrogating pulse are created from different sizes of scatterers. Secondly, these effects can be masked by strong frequency dependent attenuation. However, after careful compensation for attenuation, the H-scan bank of matched filters is capable of discriminating between the shifts in spectral characteristics.

## II. THEORY

### A. The classes of scattering structures

Considering the many different theories of scattering from inhomogeneities, and then all of the different types of scatterers and reflectors within tissues, we seek a coherent framework for characterizing the nature of tissue scattering in order to apply sensitive matched filters to distinguish between the classes. As an overview, it should be noted that the exact solutions for strong scatterers such as spheres and cylinders can be very complex, requiring infinite series of spherical or cylindrical Green's functions (Faran Jr., 1951; Born and Wolf, 1980). The situation becomes more tractable in the case of weak scattering using the Born approximation (Rayleigh, 1918; Morse and Ingard, 1987; Insana *et al.*, 1990; Insana and Brown, 1993), where closed form solutions in terms for scattered waves can be formulated as a type of a spatial Fourier transform of the scatterer shape, or else in the case of random structures, as a transform of the correlation function (Debye and Bueche, 1949). Another useful simplified framework comes from viewing the pulse echo process as a convolution (Macovski, 1983). Within the convolution framework and assuming generic bandpass pulses, it can be shown that the transfer function of structures or scatterers can sometimes be viewed as simple frequency domain functions corresponding to a perfect plane reflector (all pass filter or transfer function of $f^0$), or a derivative ($f^1$), or in-between. For example, the ensemble average amplitude from the liver has been estimated as $f^{0.7}$ (Campbell and Waag, 1984; Parker *et al.*, 2019; Parker, 2019b), consistent with the expected scattering from the fractal branching vasculature. Transfer functions can increase as $f^2$ for Rayleigh scatterers or even higher for Hermite scatterers (Parker, 2017; Astheimer and Parker, 2018). Schematically, this framework of scattering from tissue is depicted in **Figure 1** and some specific details will be explored in the following sections.

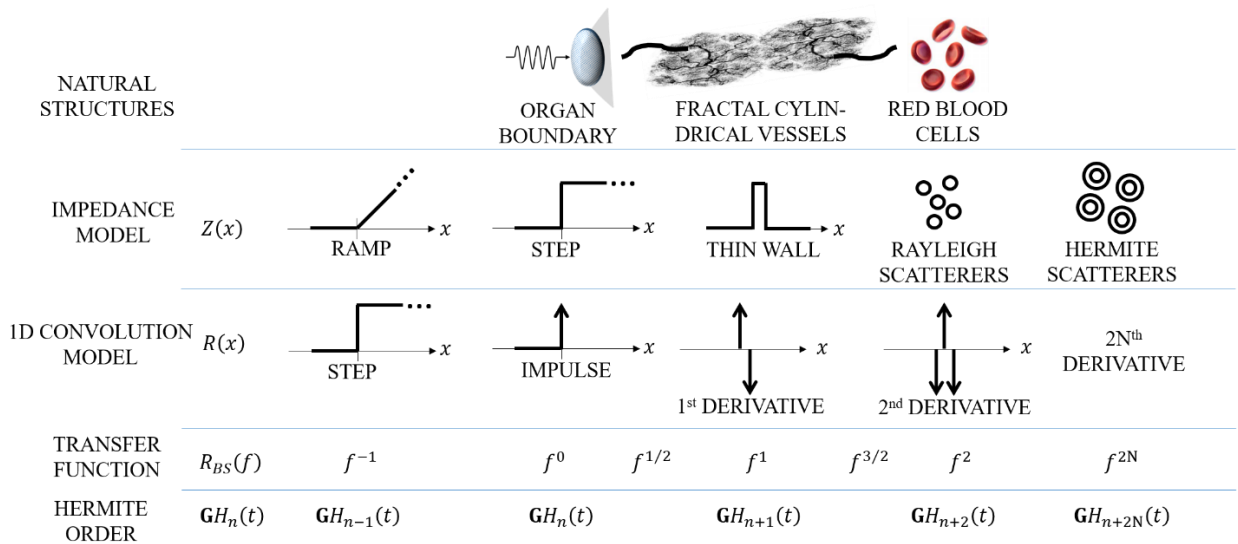

Figure 1. Schematic of shapes of inhomogeneities that cause reflections, with specific transfer functions in the frequency domain, creating specific orders of Gaussian weighted Hermite functions as echoes. Right to left is ordered in increasing power of the frequency domain reflection function. <u>Top row</u>: Acoustic impedance $z(x)$ shows a ramp function, a step function, a thin wall, Rayleigh scatterers, and Hermite scatterers. <u>Second row</u>: the corresponding reflections under the convolution model, obtained by spatial derivative of the impedance function; a step (corresponding to an integration function), an impulse, a doublet (corresponding to a first

derivative), a second derivative, and higher order derivatives. Note the addition of a random fractal ensemble of cylindrical vessels that may characterize tissue scattering sites. These have been recently found to correspond to fractional orders above or below unity. <u>Third row</u>: frequency domain transfer function corresponding to integration $f^{-1}$, all pass $f^{0}$, derivative $f^{1}$, second derivative $f^{2}$, and higher order derivatives $f^{2N}$. <u>Bottom row:</u> Matched filter order from the Gaussian weighted Hermite family corresponding to each corresponding structure, assuming a round trip impulse response of the transmitted pulse as a $GH_n(t)$ function. Most natural structures within tissues are thought to be within the $f^{0}$ to $f^{2}$ range, thus this range represents the core region for fine tuning of the H-scan analysis.

## B. Convolution models

Assume a broadband pulse propagating in the $x$ direction is given by separable functions:

$$P\left(y, z, t - \frac{x}{c}\right) = G_y(y, \sigma_y) G_z(z, \sigma_z) P_x\left(t - \frac{x}{c}\right), \tag{1}$$

where $G_y(y, \sigma_y) = \exp\left[-y^2 / 2\sigma_y^2\right]$, i.e., Gaussian in the $y$ and similarly in the $z$ directions, and where the pulse shape $P_x$ in the $x$ direction is given by:

$$P_x(x) = GH_4\left(\frac{x}{\sigma_x}\right) \exp\left(-\frac{x}{\sigma_x}\right)^2 = 4e^{-x^2/\sigma_x^2}\left[3 - 12\left(\frac{x}{\sigma_x}\right)^2 + 4\left(\frac{x}{\sigma_x}\right)^4\right], \tag{2}$$

where $GH_4$ is a fourth-order Hermite polynomial for the pulse shape with a spatial scale factor of $\sigma_x$ (Parker, 2016a; Poularikas, 2010), representing a broadband pulse. Its spatial Fourier transform is then:

$$^{3D}\Im\{P(x, y, z)\} = \left(16 e^{-k_x^2 \pi^2 \sigma_x^2} k_x^4 \pi^{9/2} \sigma_x^5\right)\left(e^{-2k_y^2 \pi^2 \sigma_y^2} \sqrt{2\pi} \sigma_y\right)\left(e^{-2k_z^2 \pi^2 \sigma_z^2} \sqrt{2\pi} \sigma_z\right), \tag{3}$$

where we use Bracewell's convention (1965) for the form of the Fourier transform. Using a 3D convolution model (Macovski, 1983; Prince and Links, 2015; Bamber and Dickinson, 1980), we will determine the *dominant* echoes.

In the alignment where the pulse is propagating perpendicular to the long axis of a cylinder, the 3D convolution result is given by the product of the transforms:

$$^{3D}\Im\{echo(x,y,z)\} = \Im^{3D}\{p(x,y,z)\}\bullet(k_x)^2\,\Im^{3D}\{cylinder(x,y,z)\} =$$
$$\left(\left(16e^{-k_x^2\pi^2\sigma_x^2}k_x^4\pi^{9/2}\sigma_x^5\right)\left(e^{-2k_y^2\pi^2\sigma_y^2}\sqrt{2\pi}\sigma_y\right)\left(e^{-2k_z^2\pi^2\sigma_z^2}\sqrt{2\pi}\sigma_z\right)\right)\bullet \quad (4)$$
$$(k_x)^2\,^{3D}\Im\{cylinder(x,y,z)\},$$

where the $(k_x)^2$ term pre-multiplying the cylinder transform stems from the Laplacian spatial derivative in the Born scattering formulation (Rayleigh, 1918; Morse and Ingard, 1987) and in the 3D convolution model (Gore and Leeman, 1977; Bamber and Dickinson, 1980). One can consider the last two terms of eqn (4) as the frequency domain transfer function which can change the shape of the interrogating pulse. This is examined further in the next section.

## C. The scattering transfer function

Consider an ideal cylindrical scatterer as a model of fluid-filled vessels, from very small microchannels of fluid to the largest arteries and veins that can exist within the organ. Specifically, we will examine a long fluid-filled cylinder of radius $a$:

$$f(r) = \begin{cases} \kappa_0 & r \leq a \\ 0 & r > a \end{cases}$$
$$F(\rho) = \frac{\kappa_0 \cdot a \cdot J_1[2\pi a \cdot \rho]}{\rho}, \quad (5)$$

where $\kappa_0$ is the fractional variation in compressibility, assumed to be $\ll 1$ consistent with the Born formulation, $F(\rho)$ represents the Hankel transform, which is the 2D Fourier transform of a radially symmetric function, $J_1[\cdot]$ is a Bessel function of order 1, and $\rho$ is the spatial frequency.

The fractional variation in compressibility, $\kappa_0$, between blood vessels and liver parenchyma has been estimated to be approximately 0.03, or a 3% difference based on published data (Parker, 2019b). This cylinder is shown in **Figure 2(a),** and the convolution transforms of eqn (4) are shown in **Figure 2(b)** and **Figure 2(c)** for the case of a pulse at normal incidence to the cylinder radius.

In addition, we consider a "soft-walled" cylindrical vessel:

$$f(r) = \frac{\kappa_0 e^{-2\pi\sqrt{\left(\frac{r}{a}\right)^2 + 1}}}{\sqrt{\left(\frac{r}{a}\right)^2 + 1}} \text{ for } r > 0 \text{ and } a > 0. \tag{6}$$

Its Hankel transform is given by theorem 8.2.24 of Erdélyi and Bateman (1954):

$$F(\rho) = \frac{\kappa_0 a^2 e^{-2\pi\sqrt{(a\rho)^2 + 1}}}{2\pi\sqrt{(a\rho)^2 + 1}} \text{ for } \rho > 0 \text{ and } a > 0. \tag{7}$$

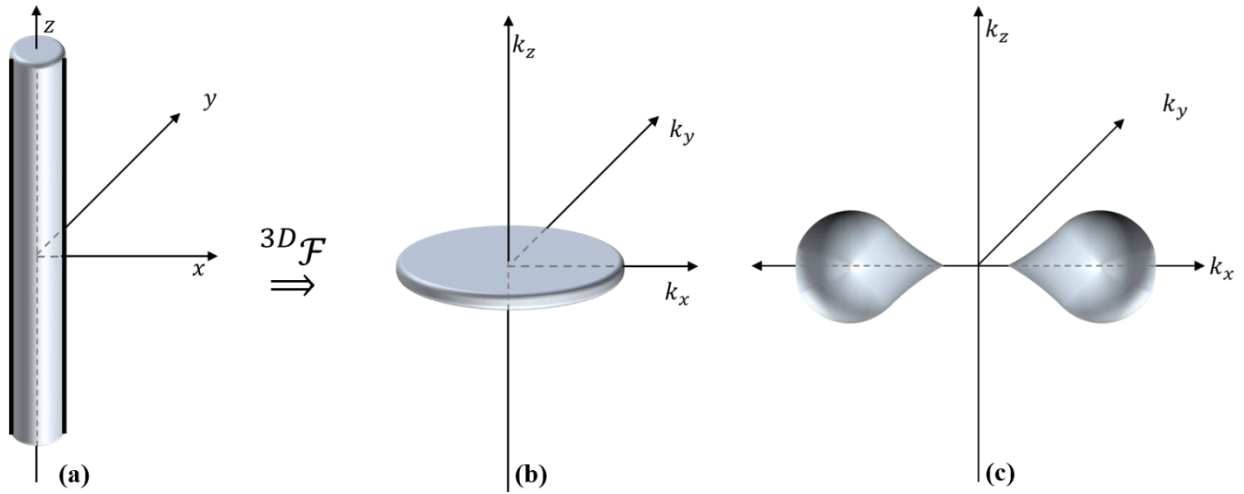

**Figure 2. (a) A cylindrical weakly scattering element. Its spatial Fourier transform is depicted in (b). The 3D spatial Fourier transform of a bandpass pulse incident normal to the cylinder is shown in (c)**

The cylindrical inhomogeneity functions of eqn (5) and eqn (6) are shown in **Figures 3(a)** and **3(b)**, respectively. When we take $\rho = \sqrt{k_x^2 + k_y^2}$ and evaluate $(k_x^2)(F(\rho))$ along the $k_x$ axis, as in in the convolution model, the result is shown in **Figure 3(b)** and **Figure 3(d)** for the two models.

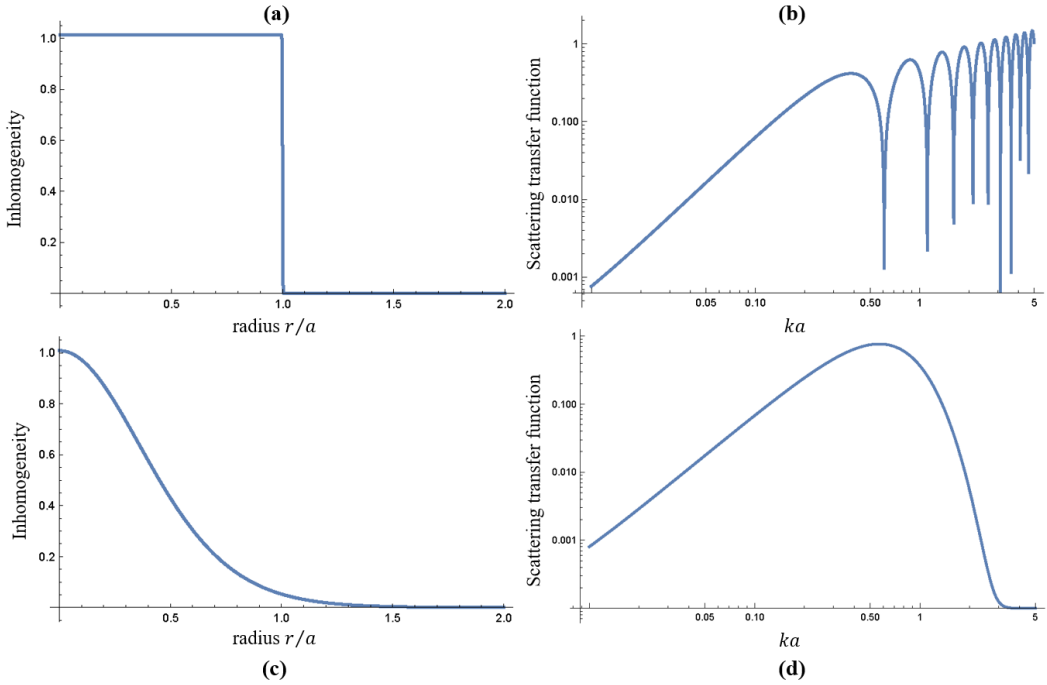

**Figure 3.** (a) Ideal cylindrical inhomogeneity function $\kappa$, and (b) its scattering transfer function in log-log scales. (c) Weak scattering inhomogeneity function $\kappa$ as a function of cylindrical radius $r$ for a "soft-walled" cylinder with unity radius. This "soft" shape avoids modeling perfect abrupt edges which may not be realized in all fluid vessels of soft tissues. (d) Log-log plot of scattering transfer function vs. wavenumber × $a$ for a "soft" cylindrical vessel at normal incidence to the propagating transmit pulse. For any narrowband pulse, the transfer function can be approximated as a power law (a slope on the log-log plot) which can vary as $f^2$ (low wavenumber, long wavelengths) to $f^0$ near $ka = 1$, to negative power law values at larger wavenumbers. These correspond to the middle ranges of transfer functions indicated in Figure 1.

Any narrowband interrogating pulse will be affected by an approximate power law transfer function, from $f^2$ at low frequencies to $f^0$ near $ka = 1$. For the ideal cylinder (**Figures 3(a-b)**), at higher $ka$, the transfer function maxima increase $\simeq f^{1/2}$. However, for the soft-walled structure, the higher $ka$ transfer function decreases as $f^{-1}$ or even more negative power laws.

## D. Generalization of backscatter for Hermite transmit functions; non-integer power laws

We assume a broadband transducer is used to produce a pulse with a round-trip impulse response that is approximated by a Gaussian-weighted Hermite function, in this case of order 4, which we designate as $GH_4(t/\tau)$, where $\tau$ is a scale factor:

$$p(t) = 4\mathrm{e}^{-\frac{t^2}{\tau^2}}\left(4\left(\frac{t}{\tau}\right)^4 - 12\left(\frac{t}{\tau}\right)^2 + 3\right), \tag{8}$$

and its Fourier transform is:

$$\Im\{p(t)\} = 4\mathrm{e}^{-(f\pi\tau)^2} f^4 \pi^{9/2} \tau^5. \tag{9}$$

The spectral peak occurs at $f_{max} = \sqrt{2}/\pi\tau$. The time and frequency domain representations of this function are shown in **Figures 4(a) and (b)**, respectively.

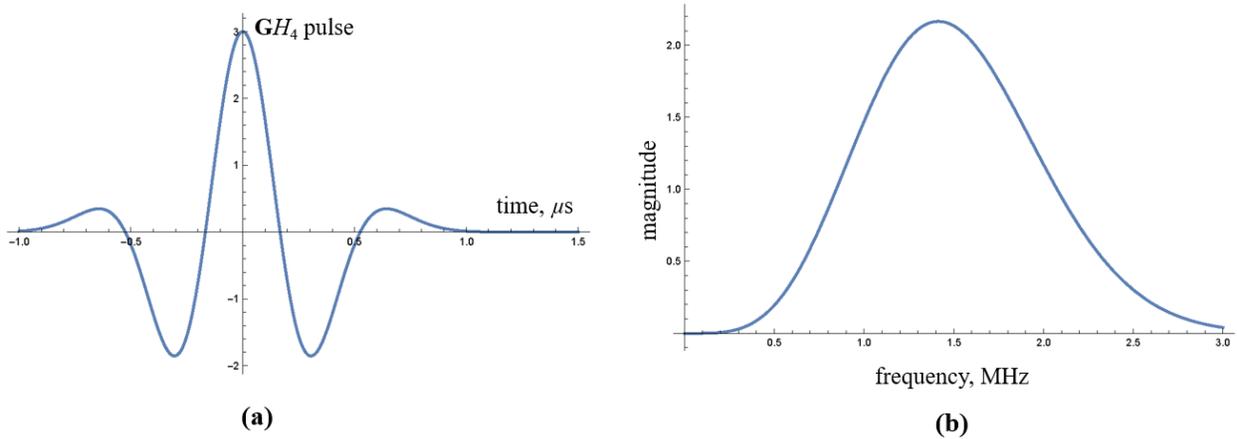

**Figure 4.** (a) $GH_4(t/\tau)$ round-trip pulse shape where $\tau$ is $1/\pi$ $\mu$s. Its Fourier transform is shown in (b) as a function of $f$. This time-frequency model is specific to the $GH$ family, but is reasonable representative of a wider class of broadband ultrasound pulses.

Now we assume the convolution model with the transfer function from scatterers of $|f|^\gamma$, where for ensemble averages over liver, we expect $\gamma$ to be near 0.7 (Campbell and Waag, 1984).

Note also that our choice of the transfer function as a real and even function of frequency implies that a real and even transmit pulse like the $\mathbf{G}H_4$ will, upon convolution with the scatterer, return an echo that remains real and even in form. In the frequency domain, the transform of the echo is given by the product of the pulse and the transfer function, thus

$$\Im\{p(t)\}\cdot|f|^{\gamma} = 4f^{(4+\gamma)}\mathbf{e}^{-(f\pi\tau)^2}\pi^{9/2}\tau^5 \quad \text{for } f > 0. \tag{10}$$

The inverse transform of this yields:

$$\mathbf{e}(t) = \frac{4}{\pi^{(\gamma+1/2)}}\left(\frac{1}{\tau}\right)^{\gamma}\text{Gamma}\left(\frac{5+\gamma}{2}\right)\text{Hypergeometric1F1}\left(\frac{5+\gamma}{2},\frac{1}{2},-\frac{t^2}{\tau^2}\right), \tag{11}$$

where the hypergeometric function is defined in chapter 15 of Abramowitz and Stegun (1964). This function has the bandpass appearance similar to that shown in **Figure 4**, but with an upshifted frequency content for positive $\gamma$.

It can be shown that for the echo spectrum of eqn (10), the peak frequency $f_{max}$ is given by:

$$f_{max} = \frac{\sqrt{(4+\gamma)}}{\sqrt{2}\pi\tau}. \tag{12}$$

This produces a relatively subtle shift in peak frequencies when tissue structures produce limited variations in $\gamma$ around 0.7 because the square root of the $4+\gamma$ term in the numerator governs this effect, thus the fine tuning of the H-scan or matched filter operations requires examinations of these incremental shifts.

We note that similar effects are produced when alternative pulses are transmitted. Although the $\mathbf{G}H_n$ pulses have general bandpass properties, some other specific models have been frequently used in simulations. For example, if we use a sine-modulated Gaussian pulse, then:

$$p_g(t) = \exp\left(-\frac{t^2}{2\tau^2}\right)\sin(\omega_0 t), \tag{13}$$

and

$$P_g(f) = -i\tau\sqrt{\frac{\pi}{2}}\left[e^{-2(f-f_0)^2(\pi\tau)^2} - e^{-2(f+f_0)^2(\pi\tau)^2}\right]. \tag{14}$$

After convolution with the scatterer transfer function of $|f|^\gamma$, the return echo can also be expressed in terms of hypergeometric functions. In the frequency domain, a shift caused by the power law can be calculated. For the Gaussian-shifted spectrum defined above, after multiplication by $|f|^\gamma$ the peak is:

$$f_{max} = \frac{1}{2}\left(f_0 + \sqrt{\frac{\gamma}{\pi^2\tau^2} + f_0^2}\right). \tag{15}$$

In this case a matched filter can be approximated by shifted Gaussians, and the similarity to $\mathbf{G}H_n$ functions can be quite close. However, the $\mathbf{G}H_n$ family retains the advantage of possessing exact derivative relationships between integer orders (Poularikas, 2010).

### E. Effect of attenuation

For the functions $\mathbf{G}H_n(t)$, derived from the $n^{th}$ derivative of a Gaussian (Poularikas, 2010; Parker, 2016b), we have the general frequency domain Fourier transform representation

$$\Im\left\{\mathbf{G}H_n\left(\frac{t}{\tau}\right)\right\} = \frac{f^n \tau^{n+1} e^{-(f\pi\tau)^2}}{\sqrt{2}} \tag{16}$$

for integers $n > 0$ and $f > 0$.

The peak frequency $f_{max}$ is determined by taking the first derivative with respect to $f$ and finding its zero. The result is $f_{max} = \sqrt{n/2}/\pi\tau$ for the $\mathbf{G}H_n(t/\tau)$ family. If one of this class of pulses is used in tissue backscatter imaging with a tissue comprised of scatterers with an ensemble average spectral magnitude of $|f^\gamma|$ and an attenuation of $\mathbf{e}^{-\alpha \cdot f \cdot x}$, then the peak frequency will be influenced by their product. Again, using the first derivative to find the peak frequency of the altered spectrum results in:

$$f_{max}(\gamma,\alpha,x,\tau,n) = \frac{\sqrt{8(n+\gamma)\pi^2\tau^2 + \alpha^2 x^2} - \alpha x}{4\pi^2\tau^2} \tag{17}$$

for $\{\gamma,\alpha,x,\tau,n\} > 0$. As $\gamma \to 0$ and $\alpha x \to 0$, the result approaches the original spectral peak, $f_{max} = \sqrt{n/2}/\pi\tau$. As will be shown in the Results section, the strong cumulative effects of attenuation can mask any subtle change in scattering power law behaviors due to alterations of tissue structure.

Alternatively, if a conventional bandpass Gaussian pulse is utilized, a corresponding frequency can be calculated. For the expected spectral magnitude of $\left(A_0 \mathbf{e}^{-(f-f_0)^2/2\sigma_f^2}\right)|f^\gamma|\left(\mathbf{e}^{-\alpha \cdot f \cdot x}\right)$, where $f > 0$, the peak frequency is given by

$$f_{max}(f_0,\sigma_f,\gamma,\alpha,x) = \frac{1}{2}\left(f_0 - \alpha\sigma_f^2 x + \sqrt{4\gamma\sigma_f^2 + (f_0 - \alpha\sigma_f^2 x)^2}\right). \tag{18}$$

These functions display a nearly linear downshift in peak frequency with depth for typical parameters found in human abdominal ultrasound. In eqn (18) $f_0$ is determined by the round-trip impulse response and results from the scatterers within tissue. In some cases, there is no reference impulse response available and average spectra are sampled from a region of interest within tissue, and then curve-fit to a Gaussian spectrum. In that case, we can re-interpret eqn (18) with $f_{max}$ as the observed ensemble average peak and $\gamma$ implicitly as 0, thus the equation reduces to

$$f_{max}(f_0, \sigma, x) = f_0 - \alpha \sigma_f^2 x, \tag{19}$$

or as an estimate from plotting the data:

$$\hat{\alpha} = \left(\frac{\Delta f_{max}}{\Delta x}\right) \cdot \left(-\frac{1}{\sigma_f^2}\right) \tag{20}$$

using the slope of $f_{max}$ versus depth. A linear downshift was first derived for attenuating media and a Gaussian pulse by (Kuc, 1984). For completeness, we note there are many additional estimators for attenuation (Parker, 1983; Parker and Waag, 1983; Cloostermans and Thijssen, 1983; Parker *et al.*, 1984; Maklad *et al.*, 1984; Wilson *et al.*, 1984; Parker and Tuthill, 1986; Parker, 1986; Taylor *et al.*, 1986; Garra *et al.*, 1987; Parker *et al.*, 1988; Zagzebski *et al.*, 1993; Girault *et al.*, 1998; Fujii *et al.*, 2002; Lee *et al.*, 2012; Tai *et al.*, 2019).

### F. Compensation for attenuation

Here we proposed an inverse filter approach to the compensation of attenuation. Once estimated by the downshift formulas, the inverse filter in the frequency domain is approximated as a real and even exponential increase as a function of frequency and depth. In practice, some upper band limit must be imposed to ensure finite energy and avoid unproductive amplification of noise. In the continuous domain, we note that a practical, finite energy inverse filter

$$\text{Infil}(\alpha, x, f, \sigma_f) = \exp(+\alpha x f) \exp\left(-\frac{f^2}{2\sigma_f^2}\right) \quad \text{for } f > 0 \tag{21}$$

will have positive amplification out to some frequency set by the $\sigma_f$ parameter, taken to be consistent with the transmit pulse's upper frequency range. If implemented as a time domain convolution, assumed to be a real and even function of time and frequency, the impulse response of this can be expressed in terms of Gaussian and erf functions, in which the complex terms appear as conjugates so the overall result is also real and even. In digital signal processing, this can be implemented as an IIR or an FIR discrete time filter of the sampled echoes.

Of course this inverse filter is depth-dependent, however the convolution for any increment of $x$ can be repeated over subsequent increments of $2x$, $3x$, $4x$, and onward since repeated convolutions in the time domain are equivalent to multiplications in the frequency domain, and $\exp(f \cdot \alpha \cdot x)^N = \exp(f \cdot \alpha \cdot N \cdot x)$ for real and positive values of the arguments.

### III. METHODS

#### A. Attenuation correction

The ultrasound signal is emitted from a transducer as shown in **Figure 5,** and we assume the RF echoes from tissue scatterers in the absence of attenuation would be $\mathrm{RF}(x)$, where $x$ is the propagated distance from the probe.

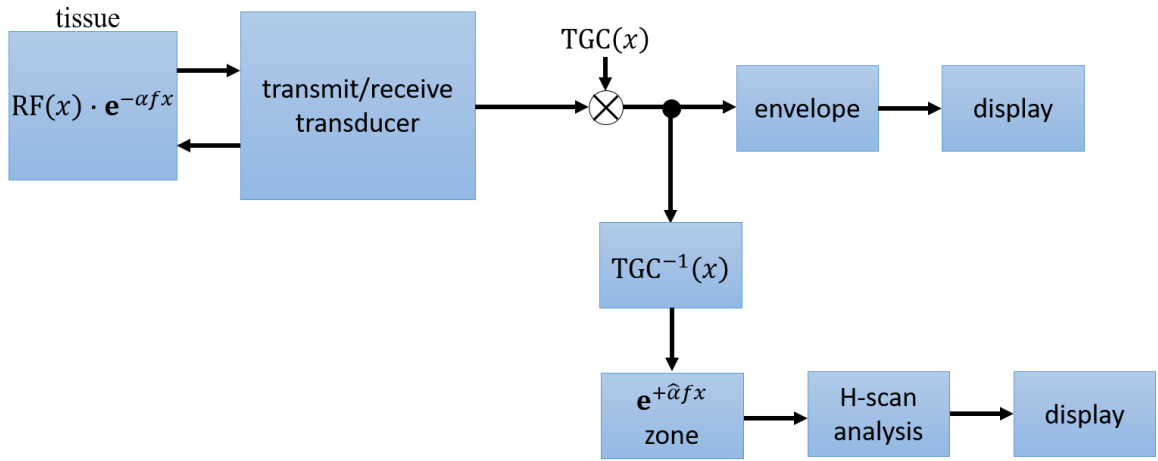

**Figure 5. Schematic for attenuation correction.**

Due to the attenuation the received ultrasound intensity can be described as $\text{RF}(x) \cdot e^{-\alpha f x}$ where $\alpha$ is the attenuation coefficient in NP/cm/MHz and $f$ is frequency of ultrasound in MHz. The received ultrasound is compensated depending on depth in the time gain compensation (TGC) amplifier, which is a practical component of ultrasound systems. To reconstruct the transmitted signal when the gain function is not recorded, we assume that the TGC was set to a function of depth that is proportional to $e^{+\hat{\alpha}_0 x}$ at the depth $x$ and $\hat{\alpha}_0$ is an *a priori* estimate of attenuation in Np/cm. Therefore, the uncompressed ultrasound signal is reconstructed as

$$\text{RS}(x) = \text{RF}(x) \cdot \text{TGC}^{-1}(x). \tag{22}$$

To compensate for frequency dependent attenuation, the transmitted signal can be reconstructed in frequency domain at some selected depth:

$$\Im\{\text{RS}(x)\} \cdot \mathcal{H}(f), \tag{23}$$

where $\mathcal{H}(f)$ is a frequency domain function to compensate for frequency-dependent attenuation. In this study, we assume $\mathcal{H}(f)$ is proportional to $e^{+\alpha_0 f x}$ within a defined region of interest (ROI) and spectral bandwidth. This is a depth-dependent filter, however to simplify the implementation, we define several zones in the depth direction. Each zone has one nominal depth, $x_z$ for each zone $z$. The reconstructed transmitted signal results in:

$$\Im\{\text{RS}(x)\} \cdot \mathcal{H}_z(f) = \Im\{\text{RS}(x)\} \cdot e^{\alpha f x_z} \tag{24}$$

for $z = \{1, 2, \ldots, n\}$, where $n$ is the number of zones. The limited dynamic range of receiver amplifiers and noise floor will always present a limit as to how far in depth the correction can be taken. Similarly, the practical upper limit of the pulse bandwidth in eqn (24) may require an upper limit to the inverse filter as given in eqn (21).

**B. Estimation of attenuation coefficient**

For the attenuation correction in Section III.A, we used one estimated attenuation coefficient for the same target when we perform multiple scans. To estimate the attenuation coefficient, the frequency spectrum can be modeled using Gaussian pulse and attenuation of ultrasound, in accord with eqn (18)–(20). To eliminate depth variable in frequency domain spectrum, this study divided ultrasound signal into several depth region, for example it can be four zones along the axial direction, as shown in **Figure 6.**

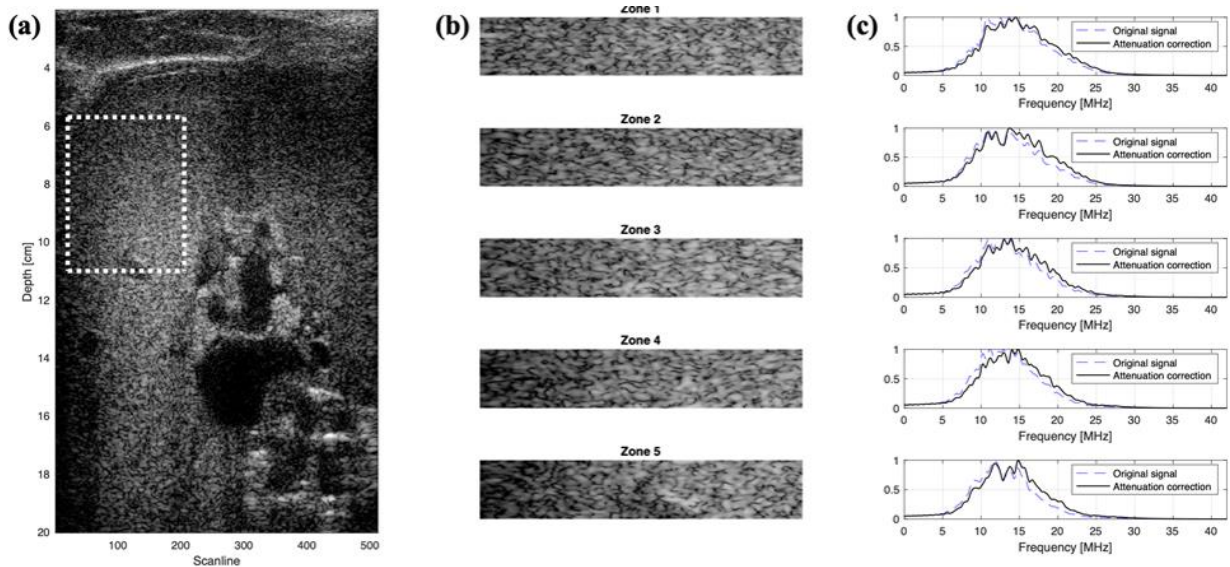

Figure 6. Attenuation-corrected spectra. Four zones.

The estimates of $\alpha_z$, the attenuation coefficient of zone $z$, is derived from the slope of $f_{max}$ versus depth. To obtain $\alpha_z$ in this equation, peak frequency for each zone is estimated using frequency spectrum, which is the Fourier transform of ultrasound signal in time domain from each zone. By selecting the maximum value of the smoothed frequency spectrum, the peak frequency for each zone can be estimated, and subsequently a linear fit of peak frequency vs. depth is calculated according to eqn (19) and eqn (20), this is shown in **Figure 7.**

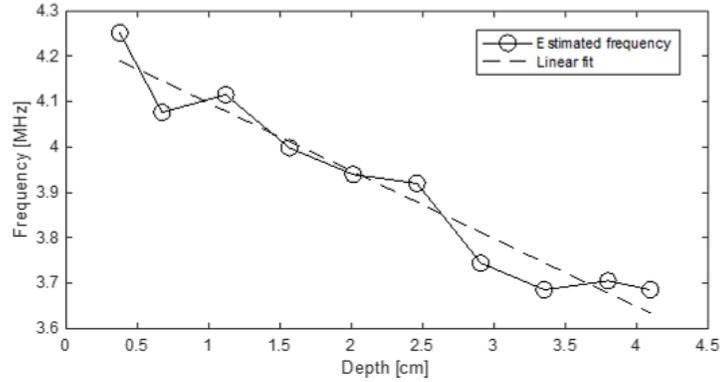

**Figure 7. Estimated peak frequencies corresponding to each zone. The linear fit of this curve is obtained, which can be used to calculate the estimated attenuation coefficient.**

## C. H-scan convolution

The H-scan concept is based on convolution with *matched filters* corresponding to specific sized scatterers. In the **Figure 8**, five different matched filter with different peak frequency were used to obtain convolution images of a phantom, using a 6.4 MHz center frequency transducer. The filters with the smallest peak frequency in **Figure 8(b)** highlighted the target strings in the phantom that is larger structures; otherwise, **Figure 8(f)** enhanced smaller sized speckles.

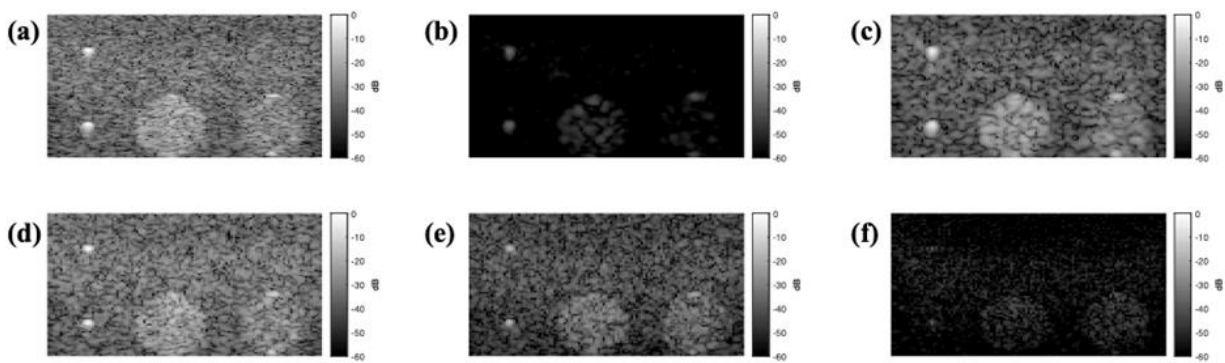

**Figure 8. (a) Input ultrasound image. (b), (c), (d), (e), and (f) are convolution images with five different matched filters, whose peak frequencies are 1.4, 3.1, 4.7, 6.3, and 7.9 MHz. The smallest peak frequency filter**

**in (b) highlights the largest structures in the phantom. The largest frequency filter of (f) tends to enhance the smallest speckle structure in the ultrasound image.**

In ultrasound images, scatterers from small to large can correspond from smaller speckles to larger structures in human body. When using different filters, various sized structures in human body can be matched to specific filters. In this study, Gaussian bandpass functions with 256 peak frequency were used to obtain convolution between the Gaussian functions and ultrasound echoes $RF(x)$. These correspond to eqn (13)-(15), where $f_0$ is taken as 4.7 MHz and $f_{max}$ ranges from 1.4 to 7.9 MHz for the case shown in **Figure 8**. The frequencies $f_0$ and $f_{max}$ are set depending on the spectrum of the input signal. Thus, each pixel in ultrasound image has a corresponding set of 256 convolution values, as shown in **Figure 9**, from which we select a maximum according to the concept of a matched filter. In our experience, these 256 outputs for each pixel have single maximum; in other words, multiple peaks are unlikely to be observed from soft tissue echoes.

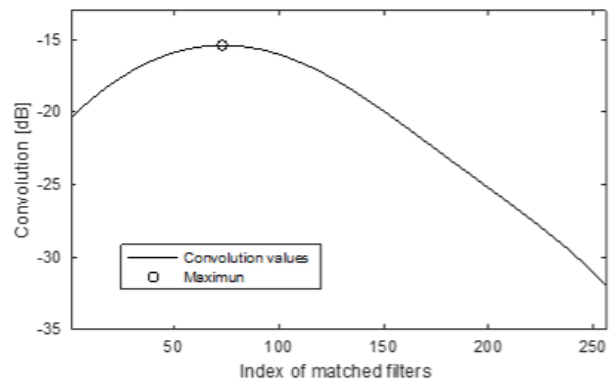

**Figure 9. Convolution values for a pixel in ultrasound image. 256 filters were used to get convolution images.**

After attenuation correction, the convolution with Gaussian functions is examined to generate H images, which are specific frequency range enhanced signals; in H1, the lowest frequency range signal is amplified, and the highest frequency is enhanced in H256. The selected indices for each pixel correspond to color map ranging from 1 to 256 RGB colors. Using transparency overlays to combine the color and B-mode data, the final H-scan image can be displayed, as described in the next section.

**D. Color map for H-scan**

To enhance visualization, a color map for H-scan is proposed, corresponding to the maximum output selected from the matched filter bank. This color map can also be thought of as a way to show the instantaneous frequency of each pixel in an ultrasound image.

A flow chart of H-scan is shown in **Figure 10**. Attenuation corrected convolution images are assigned as input data for color map processing, which generates H scan color map. By combining traditional B-scan and H-scan color map data, transparency of the two data displays H-scan image; the B-mode contains the highest spatial resolution, and the color map corresponds to peak frequency components.

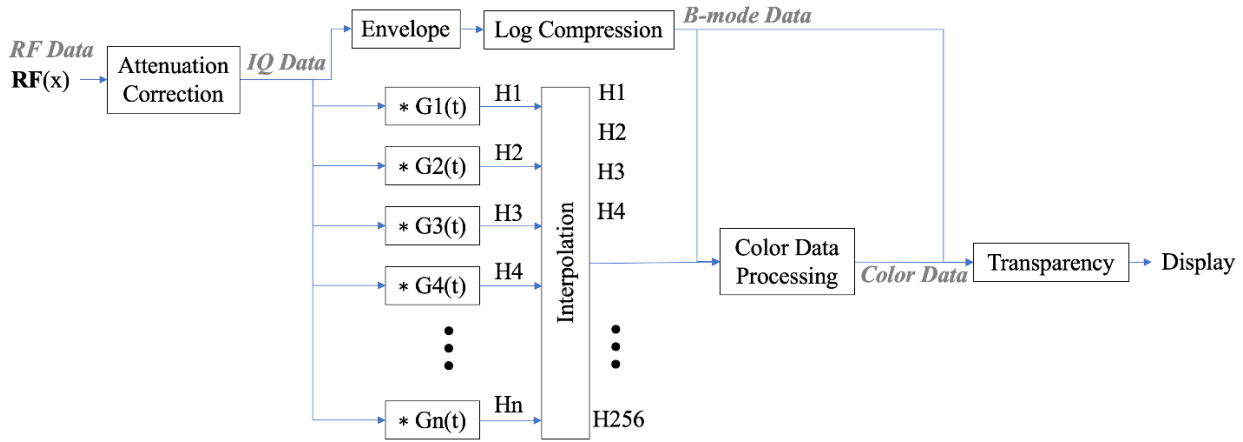

**Figure 10. Schematic for H scan.**

A color map of 256 levels is set for H scan; the levels 1 to 256 correspond from red to blue colors, sequentially, as shown in **Figure 11**. After the convolution process described in Section III.C., each pixel has one maximum matched filter output determining a specific color map level among 256, as described in **Figure 11**.

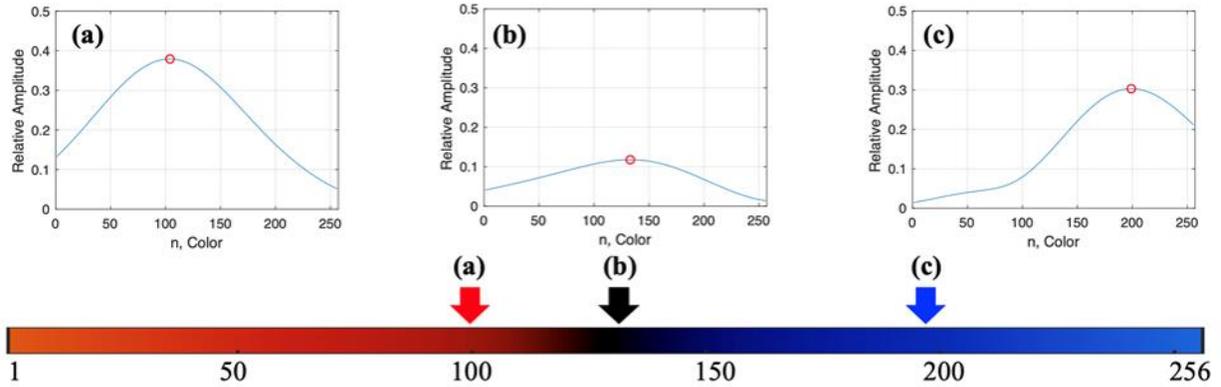

**Figure 11. For each pixel in B-mode image, there are 256 matched filter convolution values. By selecting the maximum value, we can generate a color map. (a), (b), and (c) are three pixels in a B-mode image. The *x*-axis represents indices among 256 Gaussian, and the *y*-axis represents relative amplitude of each convolution value. According to maximum amplitudes, (a), (b), and (c) pixels are assigned as red, black, and blue in the RGB color map, respectively.**

**E. Experimental setup**

H-scans with attenuation correction were obtained from *in vitro* and *in vivo* studies. To evaluate the performance of the proposed attenuation correction method, a tissue mimicking phantom (CIRS 040GSE, Computerized Imaging Reference Systems, Norfolk, VA, USA) was used. This phantom has two zones with higher and lower attenuation, whose attenuation coefficients were 0.5 and 0.7 dB/cm/MHz. For this phantom study, a Verasonics ultrasound scanner (Verasonics, Kirkland, WA, USA) with a linear array transducer (L7-4, ATL, Bothell, WA, USA) acquired RF signal at the center frequency of 5-MHz and 6.4-MHz for plane wave and focused wave, respectively. Coherent plane-wave compounding (Montaldo *et al.*, 2009) were implemented to obtain beamformed RF data.

For *in vivo studies*, rat and mouse liver data were used and there are three kinds of data: normal, hepatic fibrosis, and cancerous. For the hepatic fibrosis data, a Vevo 2100 (VisualSonics, Toronto, Canada) with a 21 MHz center frequency linear transducer (MS 250) was used to scan one normal and three fibrotic rats. Separately, a spontaneous murine model for cholangiocarcinoma (Han *et al.*, 2019) was imaged using the Vevo 3100 (VisualSonics, Toronto, CA) with a 32 MHz center frequency linear probe (MX 550D) was used. Finally, a liver metastasis model (Soares *et al.*, 2014; Ahmed *et al.*, in press), a normal and a cancerous mouse were scanned, and a Verasonics (Vantage-256, Verasonics, Kirkland, WA, USA) was used with L11-5v probe at 10 MHz center frequency. The VisualSonics scanner used focused beam transmission and the Verasonics used plane wave transmission. All *in vivo* animal studies were reviewed and approved by the University Committee on Animal Resources at the University of Rochester or the sponsoring institution.

**F. Evaluation for H-scan**

To evaluate the performance of H scan, intensity-weighted percentage of blue $(\text{IWP}_{\text{blue}})$ and red $(\text{IWP}_{\text{red}})$ were defined as follows:

$$\begin{aligned}(\text{IWP}_{\text{blue}}) &= \frac{\sum_{i \in B} I_i}{\sum_{i \in B} I_i + \sum_{i \in R} I_i} \times 100\% \\ (\text{IWP}_{\text{red}}) &= \frac{\sum_{i \in R} I_i}{\sum_{i \in B} I_i + \sum_{i \in R} I_i} \times 100\%,\end{aligned} \qquad (25)$$

where $i$ is index of each pixel in B-mode image and $I_i$ is normalized color intensity for $i$. The indices, $i \in B$ and $i \in R$, are pixels classified as blue and red, respectively.

Our color map ranges from 1 to 256; the range of 1 to 128 represents red, and that of 129 to 256 represents blue. The extreme color map indices of 1 or 256 correspond to saturated red (lowest frequency) or blue (highest frequency), respectively, so that red and blue can be normalized as $I_i$. To normalize the color data raging from 1 to 256, the red colors from 128 to 1 are re-assigned to 0 to 1 in sequence, and the blue colors from 129 to 256 are set to 0 to 1, which is given by:

$$I_i = \frac{C_i - 129}{127}, \qquad (26)$$

where $C_i$ is the color map for the $i$th pixel for blue data, and

$$I_i = \frac{128 - C_i}{127} \qquad (27)$$

for red data.

## IV. RESULTS

### A. Phantom study

In the phantom study, the attenuation correction method was evaluated using estimated peak frequency and H-scan results. Traditional B-mode image and B-mode after attenuation correction for the phantom are seen in **Figure 12(a) and (b)**, respectively. Their peak frequencies were estimated for ten depth zones in **Figure 12(d)**. B mode image has frequency down shift along depth; otherwise, the down shift after attenuation correction is compensated, resulting in more flat linear fit of estimated frequency. **Figure 12(e) and (f)** show H scan color map without and with attenuation correction, respectively. More red pixels at increasing depth are observed when attenuation correction is not used for H scan in **Figure 12(e)**. After using the attenuation correction method, red and blue pixels are more uniformly distributed over entire depth, as shown in **Figure 10(f).** The intensity-weighted percentage of blue and red were averaged for depths and it is presented as a function of depth; **Figure 10(g) and (h)** represent the percentages for the method without and with attenuation correction, respectively. In **Figure 12(g)**, red pixel percentage gradually increases, and blue pixels decrease: due to the frequency down shift. However, in **Figure 12(h)**, the percentage of blue and red pixels are similar with depth, meaning that the frequency attenuation effect is compensated by using the proposed attenuation correction method. As a result, **Figure 12(c)** demonstrates regularly-distributed red and blue colors regardless of depth in H scan image.

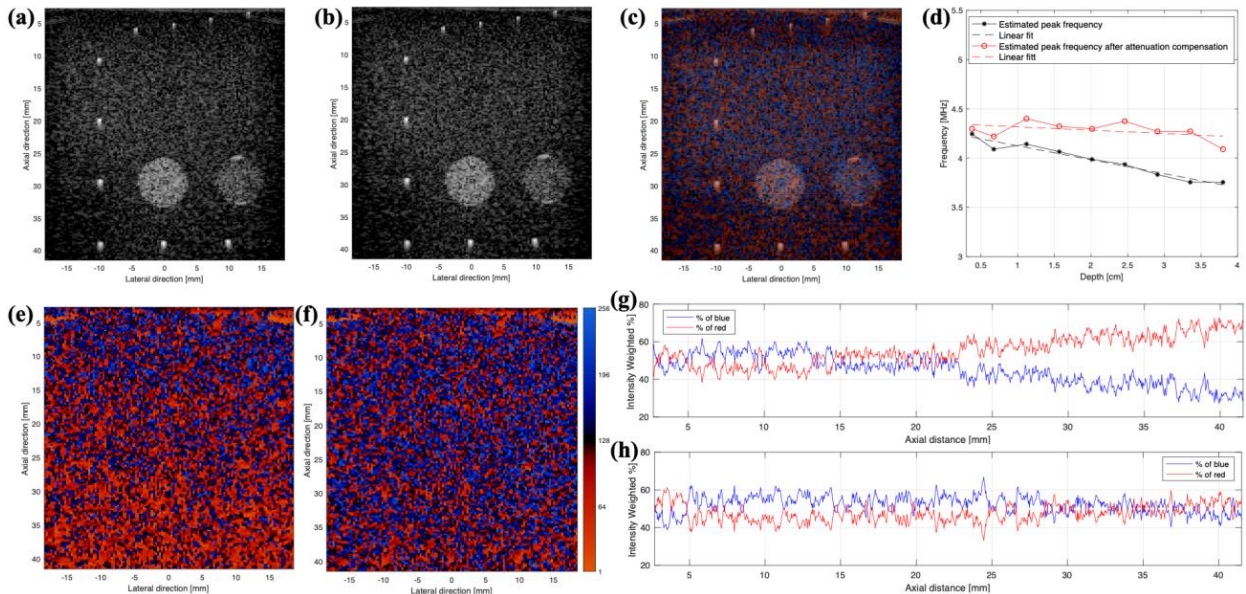

Figure 12. H scan correction for phantom: nominally 0.7 dB/MHz/cm attenuation. Verasonics L7-4, 5 MHz plane wave transmission frequency. (a) Traditional B-mode, (b) B-mode after attenuation correction, (c) H scan, (d) Estimated center frequency along imaging depth and fitting curve. (e) H scan color map for traditional RF signal, (f) H scan color map for attenuation corrected RF data, (g) Intensity weighted percentage of blue and red without attenuation correction, and (h) Intensity weighted percentage of blue and red with attenuation correction.

Other H-scan results for the phantom are shown in **Figure 13**, scanned using focused beam transmission with 6.4 MHz center frequency. Wire targets in the phantom have more red color than the background speckle since the size of the wire targets is larger than the background scatterers. To show the H-scan result more clearly, **Figure 13(d)** used threshold in color map, which shows red colored pixels with color map values ranging from 1 to 55. The clustered red colors are positioned at wire targets in the phantom.

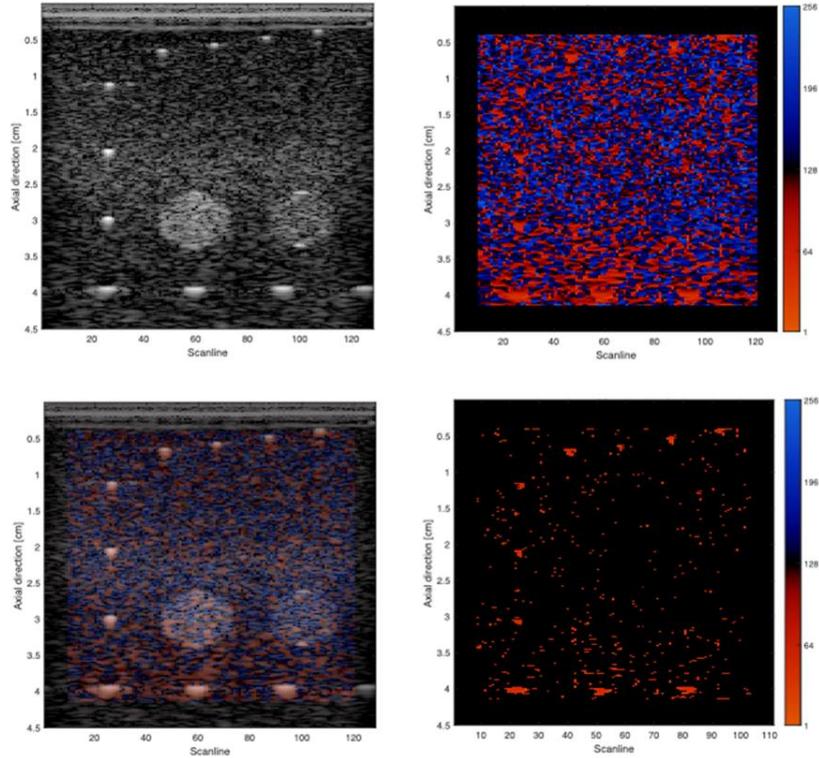

**Figure 13. B-scan and H-scan of phantom, 6.4 MHz focused transmit. (a) B-scan, (b) H-scan color map, (c) H-scan image. (d) H-scan color map with threshold to show red pixels ranging from 1 to 55 color map levels, which are corresponding to relatively larger scatterers in B scan.**

**B. Attenuation correction: *in vivo* study**

An *in vivo* scan of a rat liver with grade 4 hepatic fibrosis represents color difference between H-scan images without attenuation correction and after attenuation correction in **Figure 14**. **Figure 14(b)** shows the frequency down shift: more blue pixels in near depth, and more reds in the more distal region. However, after correction the H-scan color distribution in **Figure 14(c)** is not dependent on depth, demonstrating attenuation is compensated properly, which is consistent with phantom results. The % of blue and red profile along with the depth in **Figure 14(d)** is altered due to the attenuation effect; on the other hand, after the attenuation correction, the blue and red percentage distribution becomes more uniform as shown in **Figure 14(e)**.

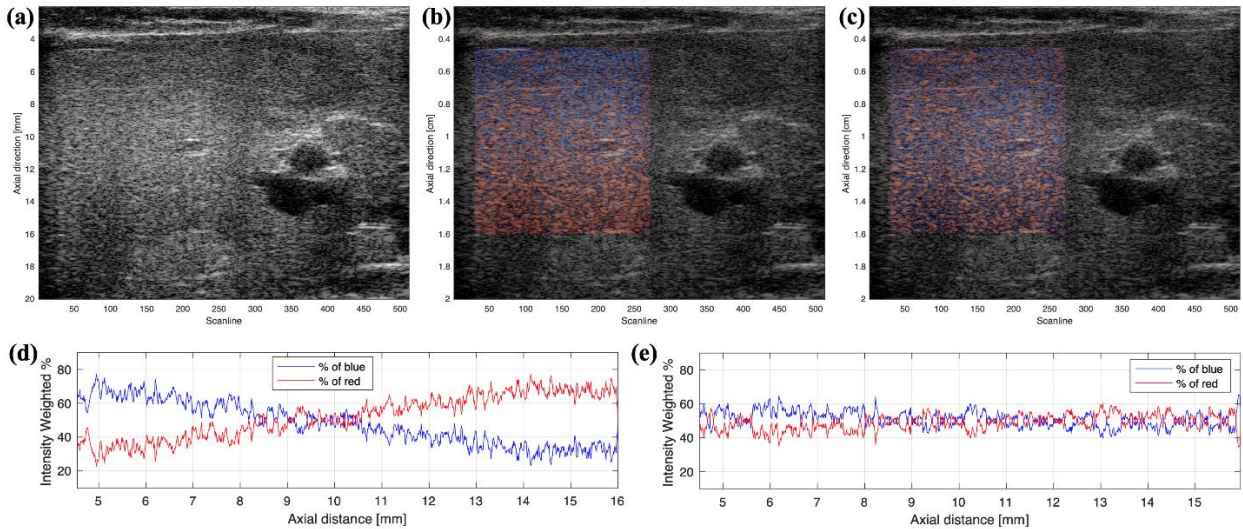

**Figure 14. Rat liver data. (a) B-mode, (b) H-scan without attenuation correction, (c) H-scan after attenuation correction, (d) intensity-weighted percentage of blue and red without attenuation correction, and (e) intensity-weighted percentages with attenuation correction.**

The attenuation correction method allows us to correct for color change caused by the frequency down shift; therefore, color difference is likely to come from variation of tissue characteristics.

In **Figure 15(b)**, a spontaneous murine tumor is poorly differentiated from normal tissue because of the shift to more red colors at depth caused by attenuation. After applying the attenuation correction shown in **Figure 15(c)**, tumor cells appear in more red than normal tissues at the same axial depth.

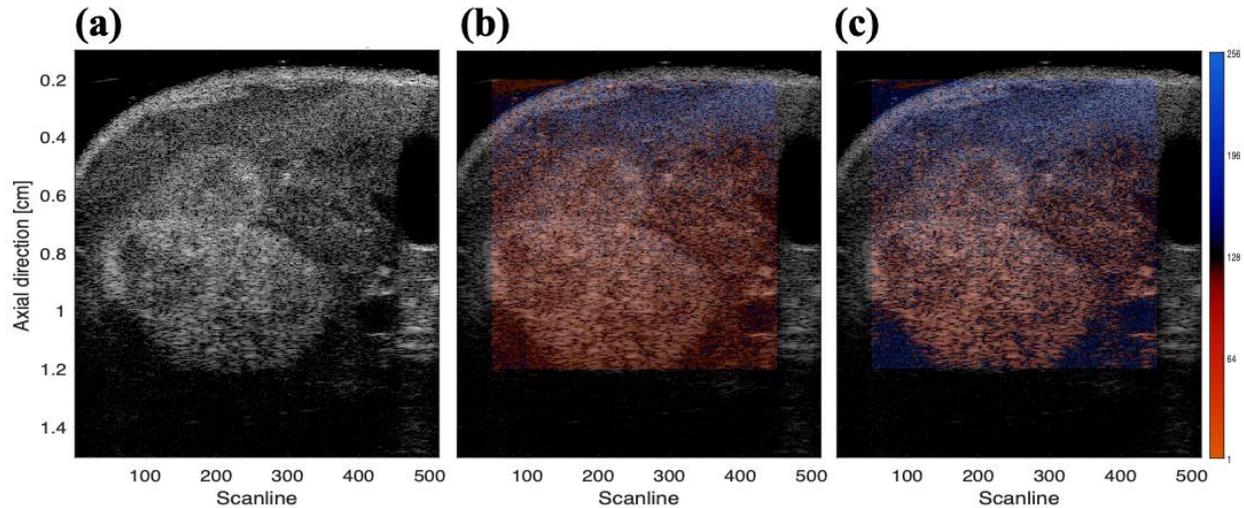

Figure 15. Murine tumor in liver. (a) B-mode, (b) H scan without attenuation correction, and (c) H scan after attenuation correction.

C. H-scan: *in vivo* study

Murine and rat *in vivo* studies with diffuse pathologies demonstrate H-scan visualizations within normal, hepatic fibrosis, and cancerous tissues. To compare hepatic fibrosis and normal liver, the H-scan ROI was set at the same depths in **Figure 16**. For the case of early and late stage tumor, livers were contoured as shown in **Figure 17**. To compare the color and percentage, the H-scan process used the same Gaussian function set for convolution in the **Figure 16** case; the **Figure 17** case also used the same Gaussian set. **Figure 16** and **Figure 17** show obvious color difference between the cases; H-scan results with hepatic fibrosis and severe stage of tumor contain more blue than the other tissues. The hepatic fibrosis case has 61.5% of the intensity-weighted blue percentage, which represents a shift to the smaller scatterers; on the other hand, the normal rat liver has 51.9%. According to the **Figure 16(c) and (f)**, the hepatic fibrosis case represents a greater difference of distribution between the red and blue pixels in H-scan. For the normalized intensity, the mean value differences between red and blue are 0.18 and 0.03 for hepatic fibrosis and normal tissue, respectively. The p-value between the red and blue scatterer groups for hepatic fibrosis is $1.0597 \times 10^{-10}$, which is smaller than the p-value of 0.1973 for normal, indicating that hepatic fibrosis has more significant difference between the red and blue groups than normal case; there are no significant differences for normal. In **Figures 16** and **17**, the following notations are used for the statistics: ns (no significance) $p > 0.05$; * $p < 0.05$; ** $p < 0.01$; *** $p < 0.001$; and **** $p < 0.0001$. In the tumor case, the smaller scatterer percentages are 50.5% for early stage (the first B-scan at 11 days after the injection of cancer cells) and 65.8% for late stage (the last scan at 28 days), and the mean value differences between red and blue are 0.01 for early stage and 0.17 for late stage. For the blue and red distribution in **Figure 17(c) and (f)**, late stage metastatic pancreatic ductal adenocarcinoma (PDAC) tumor represents greater difference between them compared to early stage; p-values between red and blue groups are 0.0813 and $1.0597 \times 10^{-10}$ for early and late stage, respectively.

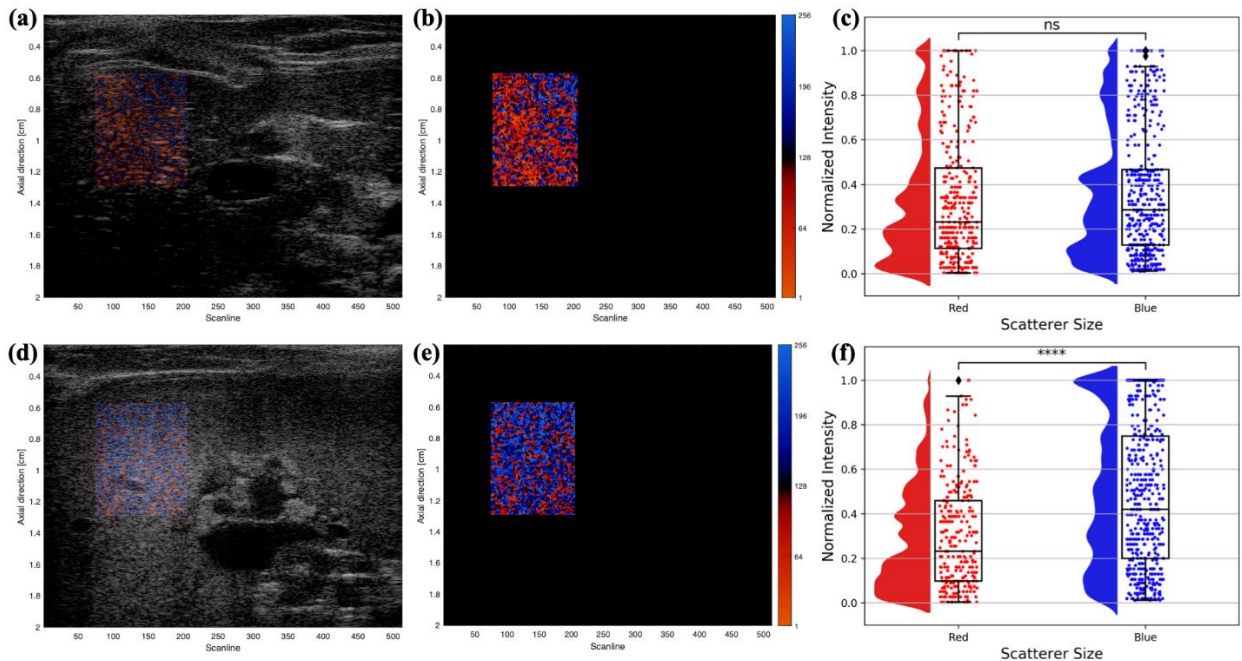

**Figure 16. H-scan of normal and abnormal rat livers. The upper row and lower row data are the results of normal and hepatic fibrosis rat, respectively. (a) and (d) show H-scan images with H-scan box, and (b) and (e) show H-scan color map data. (c) and (f) show half violin and half box plot.**

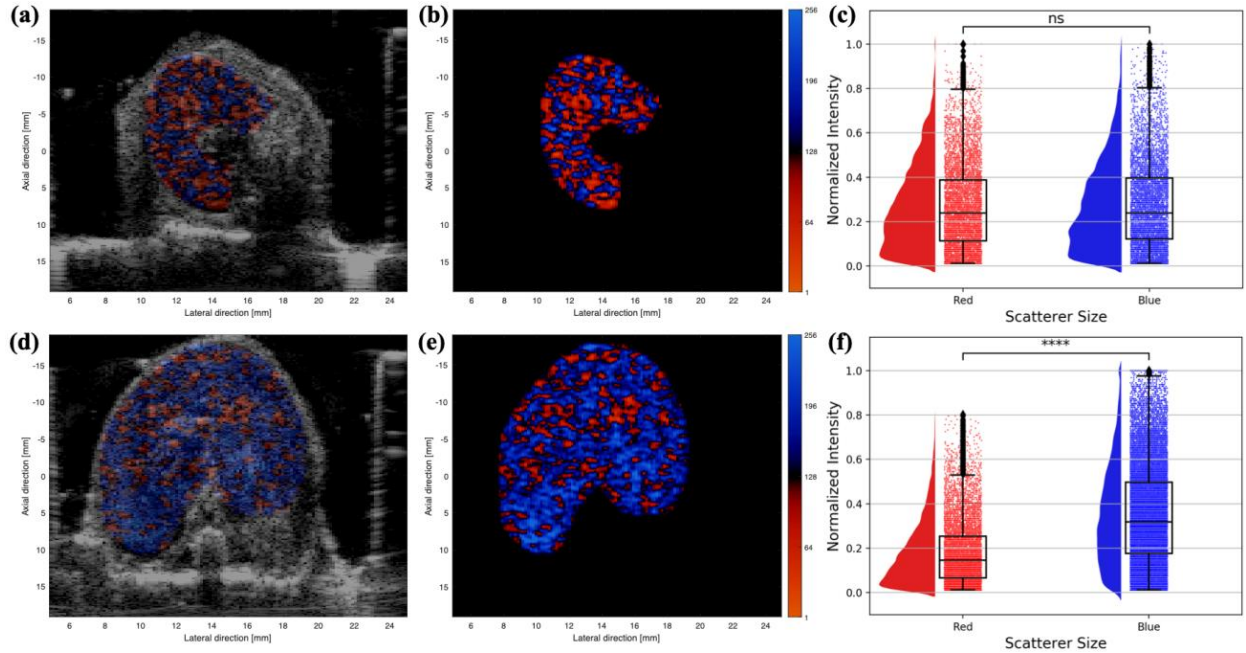

**Figure 17. H-scan of early stage and advanced metastatic cancer murine livers. Upper row data are from early stage following injection of cancer cells, and lower row data are from the liver with metastatic cancer at end stage. (a) and (d) are H-scan images. (b) and (e) are H-scan color maps. (c) and (f) show half violin and half box plot. End stage tumor tissues appear significantly more blue.**

## V. DISCUSSION

Our deterministic assessment of transfer functions from different classes of scatterers has led to a generalization in terms of a simple power law function, whereby shifts in the scattering size or type lead to shifts in the power law. Integer orders of power law 0, 1, and 2, correspond to an all-pass filter, a derivative, and a second derivative, respectively, and structures with these behaviors are easily constructed in the laboratory and may be approximated in tissues under certain conditions (Parker, 2016a). Fractional power laws above and below 1 may correspond to a physiological range of fluid filled vessels modeled as cylinders. While the exact solution for an ideal cylinder is complicated, the narrowband slope for $ka < 1$ is more easily characterized by power law behavior. This is consistent with a first order expansion, or Taylor's series expansion in log-log space around some frequency. In the long wavelength (Rayleigh scattering) regime the power law is 2, frequency squared. As $ka$ increases towards unity the slope decreases, ultimately towards zero as shown in **Figure 3(b) and (d)** for two different models of cylinders. The H-scan matched filters can be arranged so as to detect some shifts in the transfer function.

A practical question involves the minimum sensitivity of the matched filters to subtle shifts in tissue structures. The shift in peak frequency as a function of power law $\gamma$ and other factors is given by eqn (12) and (15) for the case of a $GH_4$ pulse and a Gaussian pulse, respectively. Also, the functional dependence of the power law $a$ on the cylinder radius is maximum in the long wavelength limit where transfer functions are proportional to frequency squared. Using the chain rule of calculus, for a bandpass pulse with peak spectrum at $f_{max}$, we can write:

$$\frac{\partial f_{max}}{\partial a} = \frac{\partial f_{max}}{\partial \gamma}\left(\frac{\partial \gamma}{\partial a}\right). \tag{28}$$

The first term on the right can be simply obtained for our specific cases by differentiation of either eqn (12) or (15) for $GH_4$ or Gaussian bandpass models, respectively. For example, from eqn (12) we have:

$$\frac{\partial f_{max}}{\partial \gamma} = \frac{1}{2\sqrt{2}\pi\tau\sqrt{4+\gamma}}. \tag{29}$$

The second term $(\partial \gamma/\partial a)$ is more complicated since $\gamma$ is a power law fit, but the derivative with respect to $a$ is linear. However, from **Figures 3(b) and (d)** we observe that the slope in log-log space is 2 at low $ka$ and then slowly decreases to 0 near $ka=1$. In the important subresolvable zone near $ka=0.1$ or $0.2$, we find from numerical methods that $\partial \gamma/\partial a$ is in the range of $k/4$ (units of 1/length). Using this, we have:

$$\frac{\partial f_{max}}{\partial a} \cong \frac{k}{8\sqrt{2}\pi\tau\sqrt{4+\gamma}} \quad \text{for } 0.1 < ka < 0.2. \tag{30}$$

Now examining the case where the peak frequency was 6 MHz $\cong \sqrt{2}/\pi\tau$, and $\gamma$ near 2, we have:

$$\begin{aligned}
\frac{\partial f}{\partial a} &\cong \left(\frac{k}{16}\right)(6)\left(\frac{1}{\sqrt{4+2}}\right) \\
&\cong \left(\frac{2\pi 6 \times 10^6 \text{ rad/s}}{1.5 \times 10^9 \text{ }\mu\text{m/s}}\right)(6 \text{ MHz})\left(\frac{1}{\sqrt{6}}\right) \\
&\cong 0.01 \text{ MHz}/\mu\text{m}
\end{aligned} \tag{31}$$

This result is consistent with the findings published previously (Khairalseed *et al.*, 2017), where changes in scatterer radius on the order of 10 $\mu$m were separated by a set of 4 H-scan filters, implying a 0.1 MHz discrimination in outputs.

Limitations of this study include the effect of noise on the performance of the matched filters. As can be inferred from **Figures 9, 10, and 11**, there can be substantial overlap in the spectra of neighboring matched filters, however from eqn (28) – (31), we wish to have fine grain sensitivity to changes in the scattering transfer function. Broadband noise can degrade the correct choices of a matched filter bank by adding uncorrelated and random components to the signal. These interactions and their parameterization are left for future study, however it can be concluded generally that the accuracy of selecting the correct matching filter will degrade as the signal-to-noise ratio decreases. Another limitation of this study is the reliance on power law transfer functions. As shown in **Figure 3(b)**, the transfer function from a long cylindrical vessel can be complicated at higher $ka$. However, a few factors mitigate this obvious discordance. First, from calculus, the first order approximation to a well-behaved function always exists so the concept of slope (in log-log space) for a narrowband function has meaning and justification. The question here would be how narrowband must an incident pulse be so as to make this a good approximation. In the higher $ka$ regime, the "scalloping" of the transfer function is seen in **Figure 3(b)**, however it should be noted that for an imaging pulse at higher frequencies the walls of the cylinder become resolvable, yet the interference of the front and back wall echoes combine in the Fourier transform domain to make the interference peak and valley patterns in the transfer function. This situation requires further analysis since the matched filters will at some scale be able to resolve the front and back echoes, casting a question over how high in $ka$ the transfer function of **Figure 3(b)** can be taken at face value.

## VI. CONCLUSION

The H-scan analysis implemented as a set of matched filters is capable of distinguishing relatively small shifts in scatterer transfer functions, corresponding to small changes in size of the dominant scatterers within a region. The effects of frequency-dependent attenuation can be a confounding factor at increasing depths, however a zone approach to attenuation correction can be carried out to mitigate these effects, at least to some noise-limited depth. These analyses can be cast into a deterministic framework whereby the types of scatterers encountered by an incident bandpass pulse have a scattering transfer function that is modeled as a power law. The general trend is for smaller structures to have higher power laws. Within natural structures of soft tissues, the power law $\gamma$ will frequently lie between 0 and 2, with many important changes in tissue due to pathologies resulting in subtle shifts from baseline values.

**ACKNOWLEDGEMENTS**

This work was support by National Institutes of Health grant R21EB025290. The authors also thank: Terri Swanson of Pfizer Inc. for the fibrosis data; Rifat Ahmed and Dr. Marvin Doyley of the University of Rochester Department of Electrical and Computer Engineering for the data from the metastatic tumor model; and Dr. Luis Ruffolo of the University of Rochester Medical Center Department of Surgery for data from the spontaneous murine tumor scans.